\documentclass[prb,twocolumn,showpacs,preprintnumbers,amsmath,amssymb,superscriptaddress, bibnotes]{revtex4}

\usepackage{color}
\usepackage{graphicx}% Include figure files
\usepackage{dcolumn}% Align table columns on decimal point
\usepackage{bm}% bold math

\begin{document}

\title{Phase diagram of CeRuPO under pressure investigated by $^{31}$P-NMR\\
-Comparison between CeRuPO under pressure and the Ce(Ru$_{1-x}$Fe$_{x}$)PO system -
}

\author{Shunsaku~Kitagawa}
\email{shunsaku@science.okayama-u.ac.jp}
\thanks{Present address:Department of Physics, Okayama University, Okayama 700-8530, Japan}
\author{Hisashi~Kotegawa}
\author{Hideki~Tou}
\author{Ryota~Yamauchi}
\author{Eiichi~Matsuoka}
\author{Hitoshi~Sugawara}
\affiliation{Department of Physics, Kobe University, Kobe 658-8530, Japan}

\date{\today}

\begin{abstract}
We have performed $^{31}$P-NMR measurements on single-crystalline CeRuPO under pressure in order to understand the variation in magnetic character against pressure.
The NMR spectra for $H \perp c$ and $H \parallel c$ at 2.15~GPa split below the ordered temperature, which is a microscopic evidence of the change in the magnetic ground state from the ferromagnetic (FM) state at ambient pressure to the antiferromagnetic (AFM) state under pressure.
The analysis of NMR spectra suggests that the magnetic structure in AFM state is the stripe-type AFM state with the AFM moment $m_{\rm AFM} \perp c$-axis and changes by magnetic field perpendicular to $c$-axis.
In addition, the dimensionality of magnetic correlations in the spin and the $k$ space is estimated.
We reveal that three-dimensional magnetic correlations in CeRuPO are robust against pressure, which is quite different from the suppression of the magnetic correlations along the $c$-axis by Fe substitution in Ce(Ru$_{1-x}$Fe$_{x}$)PO.
%These results suggest that the difference of phase diagram between CeRuPO under pressure and Ce(Ru$_{1-x}$Fe$_{x}$)PO originates from different efect of a tuning parameter for magnetic correlations in CeRuPO.
\end{abstract}


\newcommand{\red}[1]{\textcolor{red}{#1}}

\abovecaptionskip=-5pt
\belowcaptionskip=-10pt

\maketitle

\section{Introduction} %% No sections necessary for express letters, letters and short notes
Heavy-fermion (HF) systems, in which high temperature localized $f$ electrons behave as itinerant electrons at low temperatures, show various ground states due to the competition of two effects: the Ruderman-Kittel-Kasuya-Yosida (RKKY) interaction, which is the intersite exchange interaction between localized $f$-electron moments and induces magnetic ordering, and the Kondo interaction, which forms a nonmagnetic HF state due to screening of localized  $f$-electron moments by conduction electrons.
Crucial parameters for the two interactions are the coupling between conduction electrons and $f$-electrons, $J_{cf}$ , and the density of states around the Fermi energy $D(E_{\rm F})$; thus, the quantum critical point (QCP) in HF compounds has been presumed to be induced by tuning the strength of $J_{cf}D(E_{\rm F})$.
Experimentally, it is believed that pressure or chemical substitution is a good parameter for tunning the strength of $J_{cf}D(E_{\rm F})$ and inducing a QCP\cite{A.Schroder_Nature_2000,P.G.Pagliuso_PRB_2001,M.Ohashi_PRB_2003}.

\begin{figure}[!tb]
\vspace*{0pt}
\begin{center}
\includegraphics[width=9cm,clip]{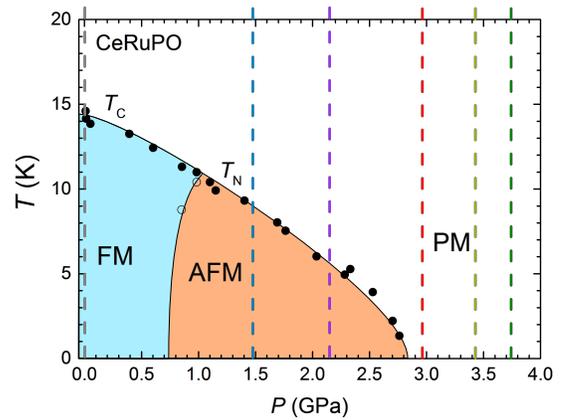}
\end{center}
\caption{(Color online) 
The pressure-temperature phase diagram for CeRuPO at zero field determined by the resistivity measurements.
%The $x$-$T$ phase diagram for Fe substitution (left side) and $P$-$T$ phase diagram (right side) for CeRuPO at zero field.
%The $x$-$T$ phase diagram is determined by $^{31}$P-NMR measurements and $P$-$T$ phase diagram is determined by resistivity measurements.
The broken lines indicates the pressure at which we have performed NMR measurements.}
\label{Fig.0}
\vspace*{-0pt}
\end{figure}

CeRuPO is a rare example of the ferromagnetic (FM) Kondo lattice among 4$f$ electron systems\cite{C.Krellner_PRB_2007}. 
Its Curie temperature $T_{\rm C}$ = 15~K and coherent temperature $T_{\rm K}$ $\simeq$ 10~K. 
It is reported that $T_{\rm C}$ is suppressed by pressure or Fe substitution at Ru site\cite{S.Kitagawa_PRL_2012,M.E.Macovei_PysicaB_2009,H.Kotegawa_JPSJ_2013}. 
We previously reported that the continuous suppression of a FM transition in Ce(Ru$_{1-x}$Fe$_{x}$)PO with an isovalent Fe substitution for Ru, and that the novel FM QCP in Ce(Ru$_{1-x}$Fe$_{x}$)PO, which is induced by tuning the dimensionality of magnetic correlations, is present at $x$ $\sim$ 0.86\cite{S.Kitagawa_PRL_2011,S.Kitagawa_PRL_2012,S.Kitagawa_JPSJ_2013}. 
The observed criticality is in sharp contrast to the FM criticality observed in other FM compounds\cite{M.Uhlarz_PRL_2004,C.Pfleiderer_PRB_1997,F.Levy_NaturePhys_2007,H.Kotegawa_JPSJ_2011,K.Adachi_JPSJ_1979,D.Aoki_JPSJ_2011}. 
For example, the FM transition in UGe$_{2}$ is gradually suppressed by applying pressure, but the transition changes from the second order to the first order at a tricritical point and the first-order metamagnetic transition emerges in a small external field\cite{H.Kotegawa_JPSJ_2011}.
On the other hands, pressure dependence of resistivity indicates a first-order transition from the FM state to another at approximately 0.7~GPa\cite{H.Kotegawa_JPSJ_2013}.
The temperature of the transition into the second phase clearly decreases under the magnetic field along the $ab$-plane. 
This is in sharp contrast to the FM state at ambient pressure, where the transition temperature increases with increasing field for both $H \perp c$ and $H \parallel c$.
The suppression of the ordered phase under the magnetic field suggests that the second phase is not an FM phase but an antiferromagnetic (AFM) phase.
The pressure ($P$)-temperature ($T$) phase diagram is shown in Fig.~\ref{Fig.0}\cite{H.Kotegawa_JPSJ_2013}. 
%In addition to the collapse of the AFM state under pressure and a magnetic field, a metamagnetic transition from a paramagnetic (PM) state to a polarized PM state appears. 
The $P$-$T$ phase diagram is quite different from the $x$-$T$ phase diagram for Fe substitution. 
Thus, it is interesting that the origin of the different phase diagrams is between the $P$-$T$ and the $x$ -$T$ phase diagram.

%%%%%%%%%%%%%%%%%%%%%%%%%%%% Figure 0 %%%%%%%%%%%%%%%%%%%%%%%%%%%%%%%%%%%%%
%%%%%%%%%%%%%%%%%%%%%%%%%%%%%%%%%%%%%%%%%%%%%%%%%%%%%%%%%%%%%%%%%%%%%%%%%%%

In this paper, we have performed $^{31}$P-NMR measurements on single-crystalline CeRuPO under pressure to identify the ground state above 0.7~GPa and to investigate the variation in magnetic properties against pressure. 
NMR spectra at 1.47 and 2.15~GPa split below the ordered temperature, which is clear evidence of the occurrence of AFM order.  
In addition, we estimate the dimensionality of magnetic correlations in the spin and the $k$ space.
The dimensionality of magnetic correlations in CeRuPO is three dimensional (3D) and does not change with pressure, which is in sharp contrast to Ce(Ru$_{1-x}$Fe$_{x}$)PO.
The difference in the phase diagram between CeRuPO under pressure and Ce(Ru$_{1-x}$Fe$_{x}$)PO might be explained by the different tuning parameters.
%In addition, the dimensionality of magnetic correlations in spin space and $k$ space is estimated from the anisotropy of Knight shift and the relationship between Knight shift and nuclear spin-lattice relaxation rate $1/T_1T$, respectively.

\section{Experimental}

Single crystalline CeRuPO was synthesized by the Sn-flux method\cite{C.Krellner_JCG_2008}.
An indenter-type pressure cell and Daphne7474 as the pressure-transmitting medium were utilized for pressure experiments\cite{T.C.Kobayashi_RSI_2007,K.Murata_RSI_2008}. 
The pressure at low temperature was estimated using a Pb manometer.
A conventional spin-echo technique was utilized for the following NMR measurement.

%%%%%%%%%%%%%%%%%%%%%%%%%%%% Figure 1 %%%%%%%%%%%%%%%%%%%%%%%%%%%%%%%%%%%%%
\begin{figure}[!tb]
\vspace*{0pt}
\begin{center}
\includegraphics[width=9cm,clip]{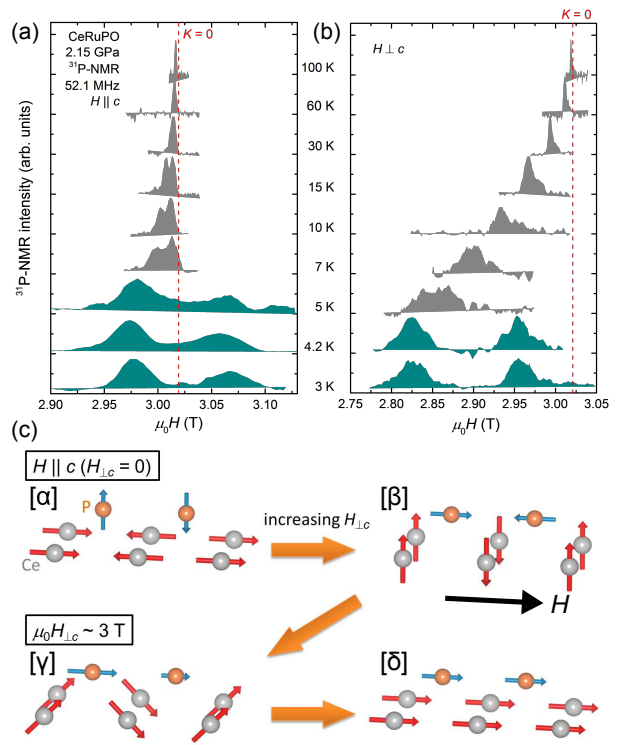}
\end{center}
\caption{(Color online) The $T$ dependence of $H$-swept $^{31}$P-NMR spectra at 52.1~MHz and at 2.15~GPa for $H \parallel c$ (a) and $H \perp c$(b).
In both magnetic field directions, NMR spectra split below the ordered temperature (representing the blue shadow), indicative of the occurrence of a two-sublattice AFM internal magnetic field at the P site.
The broken lines indicate the field at $K = 0$.
(c) The most reliable candidate of the magnetic structure based on the NMR results. 
The red arrows indicate magnetic moments of the Ce electron spins.
The blue arrows indicate internal field at the P site.
At zero field (or in $H \parallel c$), the stripe-type AFM state with $m_{\rm AFM} \perp c$ is realized [$\alpha$].
When we apply magnetic field perpendicular to $c$-axis, the spin flop occurs at a certain field and the magnetic structure becomes a stripe-type AFM state with $m_{\rm AFM} \parallel c$ [$\beta$].
The magnetic moments cant along the $ab$-plane, which is the direction of the applied magnetic field with increasing field [$\gamma$].
Finally, the magnetic structure becomes ferromagnetically at a critical magnetic field [$\delta$].}
\label{Fig.1}
\vspace*{-0pt}
\end{figure}
%%%%%%%%%%%%%%%%%%%%%%%%%%%%%%%%%%%%%%%%%%%%%%%%%%%%%%%%%%%%%%%%%%%%%%%%%%%

\section{Results and Discussion}

\begin{figure}[!tb]
\vspace*{0pt}
\begin{center}
\includegraphics[width=7cm,clip]{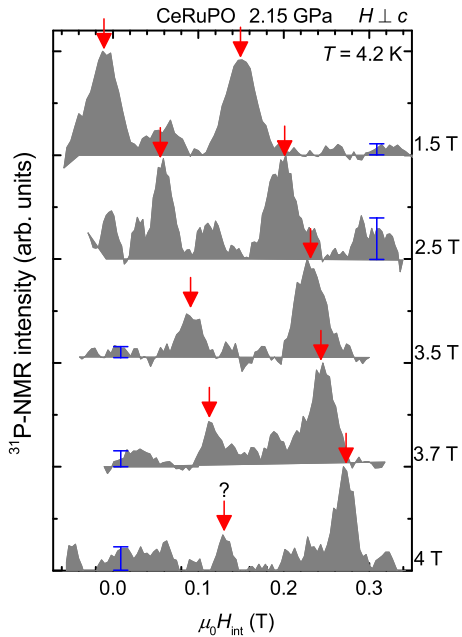}
\end{center}
\caption{(Color online) The $H$ dependence of the distribution of internal magnetic field at $T$ = 4.2~K and $P$ = 2.15~GPa deduced from the $^{31}$P-NMR spectrum.
The arrows indicate the peaks of signal.
The error bars indicate the background level.
The intensity of the peak in lower internal field decreases with increasing $H \perp c$ and becomes almost zero at 4~T.
}
\label{Fig.2b}
\vspace*{-0pt}
\end{figure}

First, we discuss the magnetic order state above 0.7~GPa, which is suggested to be an AFM state.
Figures~\ref{Fig.1} (a) and (b) show the temperature evolution of NMR spectra at 2.15~GPa in $\sim$ 3~T for $H \parallel c$ and $H \perp c$, respectively.
NMR spectra were obtained with the field-swept technique.
In $H \parallel c$, NMR spectrum shows two peaks below 20~K, suggestive of occurrence of short-range AFM order.
On cooling, the NMR spectrum significantly splits below the AFM ordering temperature $T_{\rm N}$ without a center shift, indicating that a two-sublattice AFM internal magnetic field $H_{\rm int}$ lies parallel and antiparallel to external magnetic field direction ($c$-axis) at the P site.
On the other hands, the NMR spectrum in $H \perp c$ shifts lower fields on cooling and splits below $T_{\rm N}$ with a finite center shift as shown in Fig.~\ref{Fig.1} (b).
This suggests the appearance of a two-sublattice AFM internal magnetic field with a finite FM internal magnetic field parallel to the $ab$-plane at the P site.
Comparing the $^{31}$P-NMR spectra in $H \parallel c$ and $H \perp c$, there is no magnetic structure that explains all the $^{31}$P-NMR spectra.
Then, the $H$ dependence of the magnetic structure should be considered.
The $H$ dependence of the resistivity indicates that the ordered temperature above 0.7~GPa is suppressed by $H \perp c$, although the magnetic field parallel to $c$-axis does not affect the ordered temperature\cite{H.Kotegawa_JPSJ_2013}.
Therefore, we assume that the magnetic structure changes only for $H \perp c$.
In order to check the validity of our assumption, the $H$ dependence of the $^{31}$P-NMR spectra in the ordered state was measured at $T$ = 4.2~K and $P$ = 2.15~GPa for $H \perp c$. 
As clearly seen in the figure, the two well-split NMR lines of the comparable intensity were observed below 3~T.
However, the intensity of the peak observed at the higher field side decreases with increasing $H$. 
Figure~\ref{Fig.2b} demonstrates the $H_{\perp c}$ dependence of the internal field at the P site obtained from the $^{31}$P-NMR spectra with respect to the origin of the Knight shift. 
The intensity of the peak in the lower internal field decreases with increasing $H_{\perp c}$ and disappears within the background level at 4~T, indicative of the variation in the magnetic structure.
It is considered that a spin-flop phase is reasonable since the direction of $H_{\rm int}$ changes between $H \parallel c$ and $H \perp c$.
Generally, it is difficult to realize a spin-flop phase when the spin orientation is locked by strong spin-orbit coupling.
In CeRuPO, however, the FM moments point along the magnetic field in the field, indicating that the spins can rotate in all directions\cite{C.Krellner_JCG_2008}.
Then, the spin-flop state can exist in CeRuPO.

We now consider the magnetic structure in the AFM ordered state.
Here, the P atom is located at the apex of the tetrahedra, which is formed by a P atom and four nearest neighbor Ce atoms.
This is the same configuration as the AsFe$_{4}$ tetrahedra in iron pnictides.
Therefore, the hyperfine fields at the P site $\bm{H}^{\rm P}_{\rm hf}$ can be described as, 
\begin{align}
\bm{H}_{\rm{hf}}^{\rm{P}} = \sum_{i=1}^{4} \bm{B_{i}} \cdot \bm{S}_{i}
            = \Tilde{A} \bm{S},
%\label{eq.1}
\Tilde{A} = \left(\begin{array}{ccc}
  A_{a} & C      & B_{1} \\
  C      & A_{b} & B_{2} \\
  B_{1}  & B_{2}  & A_{c} \\
  \end{array} \right),
\label{eq.2}
\end{align}
when we assume that $\bm{H}^{\rm P}_{\rm hf}$ are determined by the sum of the fields from the four nearest neighbor Ce electron spin $\bm{S}$\cite{K.Kitagawa_JPSJ_2008,S.Kitagawa_PRB_2010}.
Here, $\bm{S}_{i}$ is the Ce electron spin at the $i$th Ce site, $\bm{B_{i}}$ is the hyperfine coupling tensor between the P nucleus and $i$th Ce site, $\Tilde{A}$ is the hyperfine coupling tensor ascribed to the four nearest-neighbor Ce electron spins, $A_{i}$ is a diagonal term for the $i$ direction ($i = a, b,$ and $c$), which is related to the Knight-shift components. $B_{1\{2\}}$ components are related with the stripe ($\pi$,~0) \{(0,~$\pi$)\} AFM correlations and $C$ is related with the checkerboard ($\pi$,~$\pi$) AFM correlations in the orthorhombic notation of iron pnictides. 
%%%%%%%%%%%%%%%%%%%%%%%%%%%% Figure 2b %%%%%%%%%%%%%%%%%%%%%%%%%%%%%%%%%%%%%
%%%%%%%%%%%%%%%%%%%%%%%%%%%%%%%%%%%%%%%%%%%%%%%%%%%%%%%%%%%%%%%%%%%%%%%%%%%

%%%%%%%%%%%%%%%%%%%%%%%%%%%% Figure 2 %%%%%%%%%%%%%%%%%%%%%%%%%%%%%%%%%%%%%
\begin{figure*}[!tb]
\vspace*{0pt}
\begin{center}
\includegraphics[width=15cm,clip]{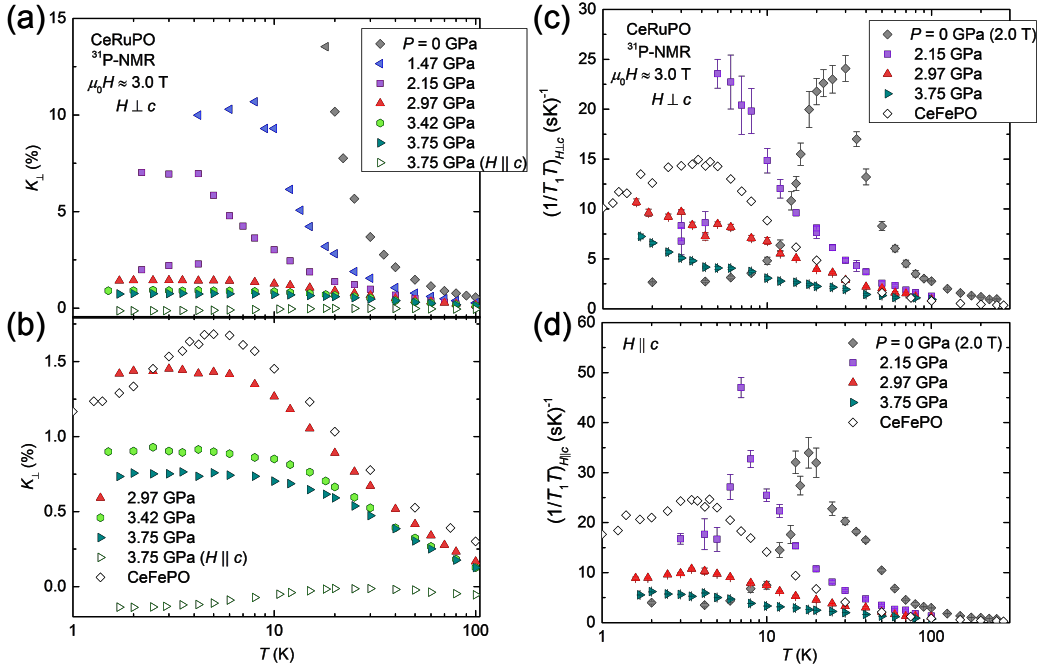}
\end{center}
\caption{(Color online) (a) The $T$ dependence of $K_{\perp}$ at various pressures and $K_{c}$ at 3.75~GPa.
The absolute value of $K_{\perp}$ at low temperature is much larger than that of $K_{c}$, indicative of the two dimensional (2D) $XY$-type anisotropy for the static spin.
$K_{\perp}$ in the PM state continuously decreases with increasing pressure.
(b) The $T$ dependence of $K_{\perp}$ above 2.97~GPa and $K_{c}$ at 3.75~GPa.
$K_{\perp}$ on CeFePO in 0.6~T is also plotted for the comparison\cite{S.Kitagawa_PRL_2011}.
The $T$ dependence of $1/T_1T$ at various pressures for $H \perp c$(c) and $H \parallel c$(d).
The data of CeFePO in 0.6~T are also plotted for the comparison\cite{S.Kitagawa_PRL_2011}.
At 1.47 and 2.15~GPa, $1/T_1T$ is enhanced toward $T_{\rm N}$ on cooling and suddenly drops at $T_{\rm N}$, suggestive of the first order phase transition.
$1/T_1T$ in the PM state continuously decreases with increasing pressure.
Above 2.97~GPa, spin fluctuations as well as $K_{\perp}$ becomes almost constant at low temperatures, indicative of the formation of the PM HF state.
}
\label{Fig.2}
\vspace*{-0pt}
\end{figure*}
%%%%%%%%%%%%%%%%%%%%%%%%%%%%%%%%%%%%%%%%%%%%%%%%%%%%%%%%%%%%%%%%%%%%%%%%%%%

First, we consider the magnetic structure in the AFM ordered state for $H \parallel c$, which is compatible with the $^{31}$P-NMR spectra in Fig.~\ref{Fig.1} (a). 
As mentioned above, we impose two experimental constraints: (i) the internal field at the P sites is parallel to the $c$-axis and (ii) the AFM state has a simple two-sublattice structure.
From eq.\eqref{eq.2}, $H_{\rm hf, c}^{\rm P} = B_{1}S_{a} + B_{2}S_{b} + A_{c}S_{c}.$
Then, only two configurations explain NMR spectra below $T_{\rm N}$:
% which generate a finite $c$-component of the internal field at P site:
(a1) the stripe-type AFM order with the AFM moments $m_{\rm AFM}$ directed perpendicular to the stripe, or (a2) the FM layers in the $ab$-plane with the FM moments directed parallel to the $c$-axis stack antiferromagnetically along the $c$-axis.
On the other hands, a magnetic structure of the AFM part in the ordered state for $\mu_0 H_{\perp c} \sim 3$~T can be presumed from Eq.\eqref{eq.2} as follows:
(b1) the stripe-type AFM order with $m_{\rm AFM} \parallel c$, (b2) the checkerboard-type AFM order with $m_{\rm AFM} \perp c$, or (b3) the FM layers in the $ab$-plane with the FM moments directed perpendicular to the $c$-axis stack antiferromagnetically along the $c$-axis.

Figure~\ref{Fig.1} (c) shows the most reasonable magnetic structures to explain all NMR spectra.
At zero field (or in $H \parallel c$), the stripe-type AFM state with $m_{\rm AFM} \perp c$ is realized [$\alpha$].
When we apply the magnetic field perpendicular to the $c$-axis, the spin flop occurs at a certain field and the magnetic structure becomes the stripe-type AFM state with $m_{\rm AFM} \parallel c$ [$\beta$].
The magnetic moments cant along the $ab$-plane, which is the direction of the applied magnetic field with increasing field [$\gamma$].
The stripe-type AFM state with canted $m_{\rm AFM} \parallel c$ can explain the NMR spectra in $\mu_0 H_{\perp c} \sim 3$~T.
Finally, the magnetic structure becomes ferromagnetically at a critical magnetic field [$\delta$].
In this scenario, the magnetic structure changes twice.
Actually, the magnetoresistance at 2.00 and 2.35~GPa below $T_{\rm N}$ shows another anomaly below a critical field, which is consistent with this scenario\cite{H.Kotegawa_JPSJ_2013}.
%In addition, NMR spectra at 1.47~GPa for $H_{\perp c} \sim 3$~T at low temperatures show no peak splitting (not shown), which is corresponding to $\delta$ state in Fig.~\ref{Fig.1} (c).
In addition, the relative intensity ratio of the peak at the lower internal field to the peak at the higher internal field decreases with
increasing $H$ and the peak at the lower internal field disappears within the background level at 4~T, although it seems that the peak at lower internal field remains a little even at 4~T, as shown in Fig.~\ref{Fig.2b}. 
On the other hand, the splitting between two peaks is independent of $H$ up to 3.7~T. 
These results suggest that the FM region increases in volume. 
Thus, the phase transition from the $\gamma$ state to the $\delta$ state is probably of the first order and the phase separation might occur at around 4~T. The complete $\delta$ state is realized in a higher field.
In order to check the validity of this magnetic structure, low-field NMR measurements in $H \perp c$ and/or the $H$ dependence of neutron diffraction measurements are effective.
In this paper, the measurement setting does not allow us low-field measurements, unfortunately.

Next, we discuss the $P$ dependence of static and dynamic magnetic properties.
Figures~\ref{Fig.2} (a) and (b) show the $T$ dependence of Knight shift at various pressures for $H \perp c$ and at 3.75~GPa for $H \parallel c$.
The Knight shift, which is the measure of the local susceptibility at the nuclear site, was determined from the peak field of each resonance spectrum.
The Knight shift $K_i$($T$, $H$) ($i = \perp$ and $c$) is defined as,  
\begin{equation}
K_i(T, H) = \left(\frac{H_0 - H_{\rm res}}{H_{\rm res}}\right)_{\omega = \omega_0} \propto \frac{M_i(T,H_{\rm res})}{H_{\rm res}},
\end{equation}
where $H_{\rm res}$ is a magnetic field at a resonance peak, and  $H_0$ and $\omega_0$ are the resonance field and frequency of a bare $^{31}$P nucleus and have the relation of $\omega_0 = \gamma_n H_0$ with the $^{31}$P-nuclear gyromagnetic ratio $\gamma_n$. 
$K_{c}$ proportional to the out-of-plane component of the susceptibility is small and almost $T$ independent, although $K_{\perp}$ proportional to the in-plane susceptibility shows the strong $T$ dependence originating from the Curie-Weiss behavior of $\chi(T)$ at high temperatures, indicative of the two-dimensional $XY$-type anisotropy. 
The $XY$-type anisotropy of static susceptibility is also observed in CeRuPO at ambient pressure\cite{S.Kitagawa_JPS_2014} and CeFePO\cite{S.Kitagawa_PRL_2011}.
Then, we focus on the $P$ dependence of $K_{\perp}$ to investigate the pressure evolution of magnetic properties.
%At and GPa, $K_{\perp}$ splits below $T_{\rm M}$, indicative of AFM order.
$K_{\perp}$ continuously decreases with increasing pressure and does not show any anomaly except for $T_{\rm N}$ in the pressure region where the AFM order appears.

NMR measurements can probe dynamic magnetic properties as well as static magnetic properties through the nuclear spin-lattice relaxation rate $1/T_1$.
In general, $1/T_1$ probes spin fluctuations perpendicular to the applied magnetic field, and thus $1/T_1$ in $H \parallel c$ and $H \perp c$ are described as,
\begin{align}
\left(\frac{1}{T_1}\right)_{H\parallel c} \equiv 2S_{\perp}, 
\left(\frac{1}{T_1}\right)_{H\perp c} \equiv S_c + S_{\perp}.
\end{align}
From two equations, we can know the low-energy $q$-summed spin fluctuation of in-plane ($S_{\perp}$) and out-of-plane ($S_c$) component, separately.
Here, we assume that the hyperfine coupling constant is isotropic and that the $H$ dependence of spin fluctuations is similar in both field direction for simplification.
%ピークがどうたらの話、1次転移
As shown in Figs.~\ref{Fig.2} (c) and (d), $1/T_1T$ is enhanced toward $T_{\rm N}$ on cooling and suddenly drops at $T_{\rm N}$ at 2.15 GPa, suggestive of a first order AFM transition.
Above 2.97~GPa, spin fluctuations as well as $K_{\perp}$ becomes almost constant at low temperatures, indicative of the formation of the PM HF state.
The phase diagram in Fig.~\ref{Fig.0} indicates that the AFM QCP is located at around 2.8~GPa.
Actually, the enhancement of the $A$ coefficient at around 2.8~GPa against pressure is observed\cite{H.Kotegawa_JPSJ_2013}.
However, $1/T_1T$ at 1.47 and 2.15~GPa, which is enhanced toward $T_{\rm N}$ on cooling, is linearly proportional to the Knight shift, as discussed later, indicative of the enhancement of the FM fluctuations.
There is no evidence that indicates the existence of the AFM quantum fluctuations in our NMR study.

Here, we discuss the origin of the difference between the $P$-$T$ phase diagram in this study and the $x$-$T$ phase diagram for the Fe-substitution system on the basis of the variation in magnetic properties.
CeRuPO shows a wide variety of ground states by applying pressure as shown above, although $T_{\rm C}$ of Ce(Ru$_{1-x}$Fe$_{x}$)PO continuously decreases with increasing $x$ toward 0~K.
In order to discuss the difference, we consider the variation in the magnetic dimensionality against the tuning parameters, which is strongly related to the phase diagram of Ce(Ru$_{1-x}$Fe$_{x}$)PO.
NMR data give us two pieces of information about the magnetic dimensionality.
One is the dimensionality in the spin space estimated from the anisotropy of $K$ and $1/T_1$.
Another is the dimensionality of the magnetic correlations estimated from the relationship between $K$ and $1/T_1T$.
We reported that the dimensionality in the spin space and the dimensionality of the magnetic correlations becomes 2D by Fe substitution in Ce(Ru$_{1-x}$Fe$_{x}$)PO\cite{S.Kitagawa_JPSJ_2013}.

%%%%%%%%%%%%%%%%%%%%%%%%%%%% Figure 3 %%%%%%%%%%%%%%%%%%%%%%%%%%%%%%%%%%%%%
\begin{figure}[!tb]
\vspace*{0pt}
\begin{center}
\includegraphics[width=8cm,clip]{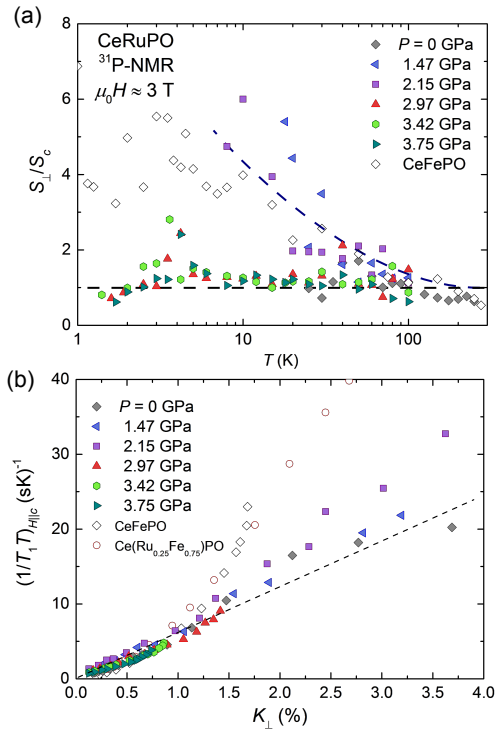}
\end{center}
\caption{(Color online) (a) The $T$ dependence of $S_{\perp}$/$S_c$ derived from $1/T_1T$ above $T_{\rm C}$ or $T_{\rm N}$ at various pressures.
The data of CeFePO are also plotted for comparison.
At 1.47 and 2.15 GPa, the $S_{\perp}$/$S_c$ increases on cooling, suggesting that the in-plane AFM fluctuations are enhanced toward $T_{\rm N}$.
At 0~GPa and above 2.97~GPa, the $S_{\perp}$/$S_c$ is $\sim$ 1 in all temperature ranges.
This result indicates that the isotropic FM fluctuations in the spin space do not change by pressure.
The broken lines are eye guides.
(b) $(1/T_1T)_{H \parallel c}$ vs $K_{\perp}$ above 10~K for various pressures.
For comparison, the data of CeFePO and Ce(Ru$_{0.25}$Fe$_{0.75}$)PO are also plotted, which have the relation of $(1/T_1T)_{H \parallel c} \propto K_{\perp}^2$ and $(1/T_1T)_{H \parallel c} \propto K_{\perp}^{1.5}$, respectively\cite{S.Kitagawa_JPSJ_2013}.
For all pressures, $(1/T_1T)_{H \parallel c}$ is linearly proportional to the $K_{\perp}$, indicating that three-dimensional FM correlations are dominant and that the dimensionality of magnetic correlations does not change against pressure.
These results indicate that pressure and Fe substitution play quite different roles for the magnetism in CeRuPO.
The broken line is the guide to eyes.}
\label{Fig.3}
\vspace*{-0pt}
\end{figure}
%%%%%%%%%%%%%%%%%%%%%%%%%%%%%%%%%%%%%%%%%%%%%%%%%%%%%%%%%%%%%%%%%%%%%%%%%%%

In CeRuPO under pressure, the $K$ in the PM state keeps the $XY$-type anisotropy as already discussed, which is similar to Ce(Ru$_{1-x}$Fe$_{x}$)PO\cite{S.Kitagawa_PRL_2012}.
As for the anisotropy of spin fluctuations, $S_{\perp}$/$S_c$ derived from $1/T_1T$ at 1.47 and 2.15~GPa increases on cooling, suggesting that the in-plane spin fluctuations are enhanced toward $T_{\rm N}$ as shown in Fig.~\ref{Fig.3} (a).
Above 2.97~GPa, the $S_{\perp}$/$S_c$ $\sim$ 1 even at low temperatures is suggestive of isotropic spin fluctuations in the spin space.
This result indicates that the isotropic spin fluctuations in the spin space do not change against pressure since the FM fluctuations at ambient pressure are isotropic.
The $S_{\perp}$/$S_c$ $\sim$ 1 at high temperatures supports the assumption of the isotropic hyperfine coupling constant.
%This behavior is quite different from that in Ce(Ru$_{1-x}$Fe$_{x}$)PO. 
Next, we discuss the magnetic-correlation character in $S_{\perp}$ in the PM state.
The magnetic-correlation characters near the magnetic instability have been interpreted using the self-consistently renormalization (SCR) theory in HF compounds as well as in itinerant magnets\cite{Moriya_SCR,T.Moriya_AP_2000}. 
In this theory, $1/T_1T$ is determined by the predominant low-energy spin fluctuation in $q$-space and the dimensionality of magnetic correlations. 
When FM correlations are dominant, $1/T_1T$ is proportional to the power law of static spin susceptibility $\chi(\bm{q} = 0)$, which is nearly equal to the Knight shift $K$ with respect to the dimensionality of the magnetic correlations: 
\begin{align}
1/T_1T &\propto 
\begin{cases}
\chi(\bm{q} = 0) \propto K \text{~(3D FM correlations),}\\
\chi^{1.5}(\bm{q} = 0) \propto K^{1.5} \text{~(2D FM correlations).}
\end{cases}
\label{eq.4}
\end{align}
Thus, the relationship between $1/T_1T$ and $K$ gives information on the $q$-space and the dimensionality of magnetic correlations. 
Although AFM order occurs above 0.7 GPa, the Korringa ratio indicates strong FM correlations even in this pressure range.
A dominant magnetic character can be inferred from the Korringa ratio given by $1/T_1TK^2$.
In the Fermi liquid state, $1/T_1TK^2$ becomes $4\pi k_{\rm B} \gamma_{n}^2/(\hbar \gamma_e^2) \equiv S_{0}$, and the magnetic correlations modify this value by introducing an enhancement factor $K(\alpha)$, given by $1/T_1TK^2 = K(\alpha)/S_{0}$.
Here, $K(\alpha) >$ 1 and $K(\alpha) <$ 1 reflect the existence of AFM and FM correlations, respectively.
For CeRuPO, $K(\alpha)$ is much smaller than 1 in all measurement pressure ranges, suggestive of the existence of strong FM correlations.
Therefore, we can estimate the dimensionality of the magnetic correlations from Eq.\eqref{eq.4}.
Figure~\ref{Fig.3} (b) shows $K_{\perp}$ versus $(1/T_1T)_{H \parallel c} \propto S_{\perp}$ for various pressures.
For all pressures, $(1/T_1T)_{H \parallel c}$ is linearly proportional to $K_{\perp}$, indicating that 3D FM correlations are robust against pressure.
Figures~\ref{Fig.3} (a) and (b) clearly show that the magnetic fluctuations in the samples possess the isotropic character in the spin and the $k$ spaces, and they do not change under pressure, which is quite different from the case of Ce(Ru$_{1-x}$Fe$_{x}$)PO.

\begin{figure}[!tb]
\vspace*{0pt}
%\hspace*{0pt}
\begin{center}
\includegraphics[width=9.5cm,clip]{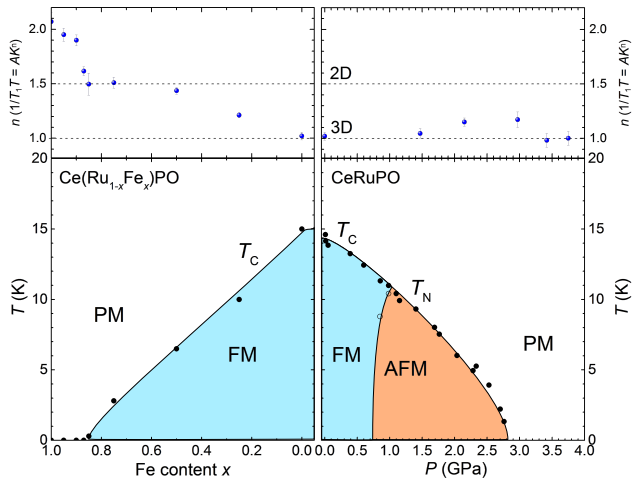}
\end{center}
\caption{(Color online) The $x$-$T$ phase diagram for Fe substitution\cite{S.Kitagawa_JPSJ_2013} (left side) and the $P$-$T$ phase diagram (right side) for CeRuPO at zero field.
The power law $n$ against $x$ or $P$ is estimated from the fitting between $\sim$$T_{\rm C}$ or $T_{\rm N}$ and 100~K and plotted in the above panel.
$n$ changes from 1.0 (3D FM correlations) to 1.5 (2D FM correlations) against $x$ although $n$ remains $\sim$ 1.0 against pressure.}
\label{Fig.6}
\vspace*{-0pt}
\end{figure}

We summarized the variation in the dimensionality of the magnetic correlations against pressure and Fe substitution together with each phase diagram in Fig.~\ref{Fig.6}.
There are mainly three features:(i) the FM QCP only appears for Fe substitution at $x \sim 0.86$, (ii) the dimensionality of magnetic correlations changes from 3D to 2D with Fe substitution while it does not change with pressure, and (iii) the AFM order is realized under pressure in spite of the strong FM correlations.
In general, chemical substitution and pressure are treated as similar tuning parameters\cite{D.Jaccard_PhysicaB_1999,H.Q.Yuan_Science_2003}.
However, sometimes quantum critical behavior is masked by chemical disorder and these two parameters play different roles for the ground state\cite{S.Seo_NaturePhys_2014}.
In the case of CeRuPO, pressure and Fe substitution are quite different tuning parameters in contrast to other systems: Fe substitution tunes the dimensionality of magnetic correlation along $c$-axis while pressure behaves like a conventional tuning parameter.
As a consequence, the two phase diagrams become different as shown in Fig.~\ref{Fig.6}.
Our NMR results suggest that a FM QCP is an energetically unstable state and that the suppression of the dimensionality of magnetic correlation along $c$-axis is important for the appearance of a FM QCP.
%These NMR results strongly indicate that pressure and Fe substitution play quite different roles for magnetism in CeRuPO, resulting that the phase diagram becomes different between the pressurized and the Fe-substitution system.

%\section{discussion}

%%%%%%%%%%%%%%%%%%%%%%%%%%%% Figure 6 %%%%%%%%%%%%%%%%%%%%%%%%%%%%%%%%%%%%%
%%%%%%%%%%%%%%%%%%%%%%%%%%%%%%%%%%%%%%%%%%%%%%%%%%%%%%%%%%%%%%%%%%%%%%%%%%%

%%%%%%%%%%%%%%%%%%%%%%%%%%%% Figure 2 %%%%%%%%%%%%%%%%%%%%%%%%%%%%%%%%%%%%%
%\begin{figure}[!tb]
%\vspace*{-0pt}
%\begin{center}
%\includegraphics[width=8.5cm,clip]{Fig2.eps}
%\end{center}
%\caption{(Color online) }
%\label{Fig.2}
%\end{figure}
%%%%%%%%%%%%%%%%%%%%%%%%%%%%%%%%%%%%%%%%%%%%%%%%%%%%%%%%%%%%%%%%%%%%%%%%%%%

\section{Summary}
In summary, we have performed $^{31}$P-NMR measurements on single-crystalline CeRuPO under pressure.
$^{31}$P-NMR results provide strong evidence of the AFM ordering at 1.47 and 2.15~GPa.
Analysis of the NMR spectra suggests that the magnetic structure in the AFM state is the stripe-type AFM state with $m_{\rm AFM} \perp c$ at zero field.
%In spite of the AFM ordering above 0.7~GPa, FM correlations are dominant in all pressure range.
In addition, the dimensionality of magnetic correlations in the spin and the $k$ spaces is estimated from the anisotropy of $K$ and $1/T_1T$, and the relationship between $K$ and $1/T_1T$, respectively.
In CeRuPO, the magnetic correlations are 3D and do not change with pressure, while the magnetic correlations become 2D with Fe substitution.
These NMR results strongly indicate that pressure and Fe substitution play quite different roles for the magnetism in CeRuPO, resulting in different phase diagrams between the pressurized and the Fe-substitution system.
Our NMR measurements demonstrate the wide variety of phase diagrams in CeRuPO.

\section*{Acknowledgments}
The authors are grateful to K. Ishida for fruitful discussions. 
This work was partially supported by Kobe Univ. LTM center,  a grant-in-aid for Scientific Research from Japan Society for Promotion of Science (JSPS), KAKENHI (B) (No. 24340085). 
One of the authors (SK) was financially supported by a JSPS Research Fellowship.

%\bibliographystyle{apsrev4-1}
%\bibliographystyle{apsrev4-1_nocomma_etal}
%\bibliography{Ref,NMR}

%

\end{document}